\documentclass{article}
\PassOptionsToPackage{numbers, compress}{natbib}
\usepackage[nonatbib, final]{neurips_2024}
\usepackage[utf8]{inputenc}
\usepackage[T1]{fontenc}
\usepackage{xurl}
\usepackage[colorlinks, allcolors=blue]{hyperref}
\usepackage{booktabs}
\usepackage{amsfonts}
\usepackage{nicefrac}
\usepackage{microtype}
\usepackage{xcolor}
\usepackage{graphicx}
\usepackage{tabularx}
\usepackage{enumitem}
\usepackage{multirow}
\usepackage{array}
\usepackage{apacite}
\usepackage[most]{tcolorbox} 
\usepackage{caption}
\definecolor{boxbackground}{HTML}{e8ebf3}
\definecolor{boxline}{HTML}{a6adc0}
\usepackage{cleveref}
\let\citep\shortcite
\let\citet\shortciteA

\title{Safety cases for frontier AI}

\author{Marie Davidsen Buhl\thanks{Corresponding author: \url{marie.buhl@governance.ai}.} \qquad
Gaurav Sett \qquad
Leonie
Koessler \\ \\
\bfseries Jonas Schuett \qquad Markus Anderljung \\\\
Centre for the Governance of AI 
}

\begin{document}
\maketitle
\setcounter{footnote}{0}

\begin{abstract}
As frontier artificial intelligence (AI) systems become more capable, it becomes more important that developers can explain why their systems are sufficiently safe. One way to do so is via safety cases: reports that make a structured argument, supported by evidence, that a system is safe enough in a given operational context.
Safety cases are already common in other safety-critical industries such as aviation and nuclear power. In this paper, we explain why they may also be a useful tool in frontier AI governance, both in industry self-regulation and government regulation. We then discuss the practicalities of safety cases, outlining how to produce a frontier AI safety case and discussing what still needs to happen before safety cases can substantially inform decisions.
\end{abstract}
\vspace{1em}

\begin{figure}[h!]
    \makebox[\textwidth][c]{\includegraphics[width=1\linewidth]{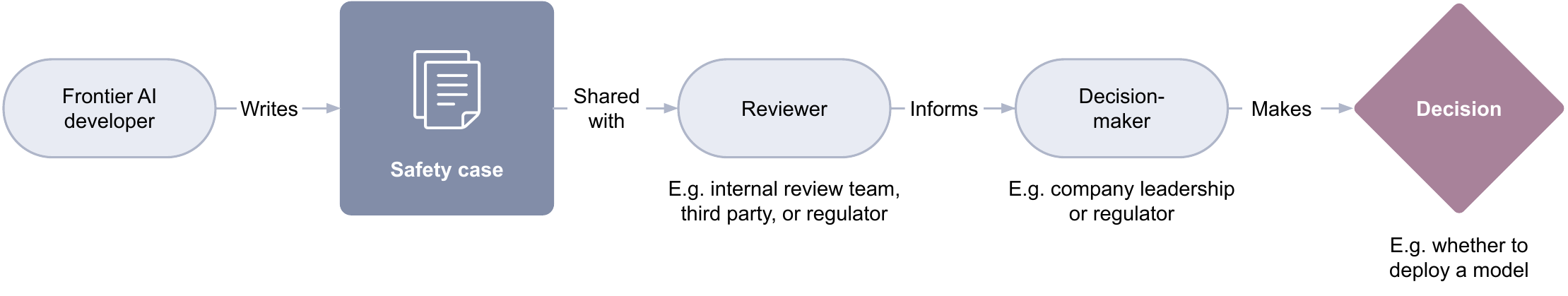}}
    \caption{Process for using a safety case to inform a decision}
    \label{fig:1}
\end{figure}

\newpage
\tableofcontents
\newpage

\section*{Executive summary}

\subsection*{What is a safety case? (\Cref{safety-cases-and-frontier-ai})}

\begin{itemize}[leftmargin=2em]
    \item A safety case is a structured argument, supported by evidence, that a system is safe enough in a given operational context. Safety cases are typically prepared by a developer ahead of a major decision (e.g. whether to deploy a system), reviewed by an independent third party (e.g. a third-party auditor), and shared with decision-makers (e.g. the board of directors or a regulator).

    \item Safety cases are common in other safety-critical industries such as nuclear power, aviation, and autonomous vehicles.

    \item Safety cases could complement existing frontier AI safety frameworks, which are organization-level policies for managing risks from frontier AI systems. By contrast, a safety case is a system-level assessment of such risks. Developers could use safety cases to explain how they have adhered to their safety framework, including its more subjective components.

\end{itemize}

\subsection*{Why use safety cases for frontier AI systems? (\Cref{the-case-for-safety-cases})}

\begin{itemize}[leftmargin=2em]
    \item Safety cases could be used internally by developers to inform high-stakes decisions (e.g. whether to deploy a system), as well as by regulators to assess compliance with safety requirements.

    \item A benefit of safety cases is that making an explicit, structured argument that a system is safe enough is a useful way to stress test risk assessments. In particular, it could help developers and third parties to spot shortcomings in developers’ risk assessments.

    \item Another benefit of safety cases is their flexibility. Safety cases allow developers to assess and mitigate risks with the methods most suitable to a particular system. This makes them a future-proof tool that will remain useful even if frontier AI continues to develop rapidly. It also capitalizes on developers’ expertise and resources.

\end{itemize} 

\subsection*{What would a frontier AI safety case look like? (\Cref{components-of-a-safety-case})}

\begin{itemize}[leftmargin=2em]
    \item A safety case has four key components: Objectives (that must be met for the system to be safe enough), arguments (that the objectives have been met), evidence (that the arguments are true), and scope (in which the safety case holds). 

    \item Safety cases for current systems could be based on the arguments implicit in existing safety frameworks. Namely, they could argue that a system is not capable enough to cause catastrophic harm by evaluating if the system has certain dangerous capabilities.

    \item For systems with significantly more dangerous capabilities, safety cases could argue that effective safeguards are in place or that the system will not use its capabilities to cause catastrophic harm.

\end{itemize}  

\subsection*{What challenges need to be addressed before frontier AI safety cases can inform decision-making? (\Cref{implementation-challenges})}

\begin{itemize}[leftmargin=2em]
    \item One challenge will be to develop methodology for frontier AI safety cases. This is still early-stage; more research and investment is needed before best practices can be established.

    \item Another challenge will be to develop adequate safety cases for future systems with more dangerous capabilities, as we do not yet know how to reliably assure the safety of such systems. Significant research into novel safety techniques is needed.

    \item A third challenge will be for developers and regulators to establish processes and build sufficient capacity to effectively review safety cases.

\end{itemize} 

\subsection*{What should be done now? (\Cref{policy-recommendations})}

\begin{itemize}[leftmargin=2em]
    \item Developers should begin to produce and share safety cases; commit to using safety cases in future deployment decisions; build up internal capacity and processes for producing and reviewing safety cases; and share lessons with other actors.

    \item Governments should encourage companies to produce and share safety cases; conduct or support relevant research; advance a third-party ecosystem; and consider using safety cases to assess compliance with potential future regulation.

\end{itemize}
\newpage

\section{Introduction}\label{introduction}

Frontier AI systems -- highly capable general-purpose AI systems that can perform a wide variety of tasks and match or exceed the capabilities present in the most advanced systems \citep{dsit2024}\footnote{But note that the term ``frontier AI'' has been criticized \citep{helfrich2024}.} -- may pose severe risk to society.\footnote{By ``risks to society'', we mean risks of significant harm to large groups of people. This does not include legal, financial, or reputation risks to frontier AI companies themselves.} Existing systems can already help conduct cyber-attacks \citep{fang2024a,fang2024,NCSC2024} and show early signs of manipulative and deceptive capabilities \citep{openai2024,park2023,scheurer2024}. Future systems may be able to assist users in producing biological weapons \citep{mouton2023,soice2023,urbina2022} or act with increasing agency, making them difficult to control \citep{chan2023,cohen2024,kinniment2024}. Eventually, AI systems may even cause catastrophic outcomes \citep{bengio2024,grace2024,hendrycks2023}. If capabilities continue to advance as rapidly as in recent years, these risks could increasingly become a reality. 

Given the potential risks, frontier AI developers\footnote{By ``frontier AI  developers'', we mean organizations that design and train complete frontier AI systems. This currently covers a handful of private companies, but could in theory also cover non-profit or governmental organizations. It does not include downstream developers who build products based on frontier AI systems.} should be able to explain why they think their systems are safe enough\footnote{We use ``the system is safe enough'' interchangeably with ``the system does not pose unacceptable risk''.} to develop\footnote{By ``development'', we mean designing, training, fine-tuning, and applying safeguards to an AI system.} or deploy.\footnote{By ``deployment'', we mean releasing a system to certain users or applying it to certain real-world tasks. There are many possible types and degrees of deployment \citep{solaiman2023}. For example, the developer may open source the model weights \citep{kapoor2024,seger2023} or only allow access via an API \citep{shevlane2024}. In writing this paper, we primarily had in mind internal deployment, although safety cases may also be used to inform internal deployment decisions.} Safety cases are one way in which developers could produce and communicate such explanations. A safety case is a structured argument, supported by evidence, that a system is safe enough to deploy in a given way \citep{stan1996}. Safety cases are used in many safety-critical industries, such as nuclear power, aviation, and autonomous vehicles \citep{cleland2012evidence,inge2007,sujan2016}. There has recently been increasing interest in using them for frontier AI as well – from academics \citep{bengio2024,clymer2024,wasil2024,yohsua2024}, governments \citep{irving2024}, and developers \citep{anthropic2024,googledeepmind2024}. However, there is still little clarity on what frontier AI safety cases would look like and how they could be used.

This paper explores the idea of using safety cases for frontier AI. Our scope is restricted in four ways. First, we focus on frontier AI systems as opposed to AI systems in general. We chose this focus because frontier AI systems are among the systems most likely to pose severe risks and present unique assurance challenges that warrant novel analysis. However, safety cases may also be a useful tool for other high-risk systems, such as biological design tools \citep{sandbrink2023} and other advanced narrow-purpose systems in high-risk domains. Indeed, safety cases are already used for autonomous vehicles \citep{favaro2023} and defense-related software \citep{stan1996}. Second, we focus on catastrophic risks\footnote{By "catastrophic risk", we mean risks of extremely large-scale harm, for example damage in the tens of thousands of lives lost, hundreds of billions of dollars of economic or environmental damage, or significant adverse disruption to the social and political order \citep{shevlane2023}.} since they are the primary focus of companies’ existing safety frameworks \citep{anthropic2024,OpenAI2023-tt,googledeepmind2024,magic2024} , though safety cases could also be used to address other risks. Third, we focus on the US, UK, and EU contexts both due to their significance in the current conversation about frontier AI governance and because they are the jurisdictions we are most familiar with. Fourth, we use deployment as our primary example, though safety cases may also be useful in informing development decisions (e.g. whether to begin a training run).

The article proceeds as follows. \Cref{safety-cases-and-frontier-ai} provides an overview of what safety cases are. \Cref{the-case-for-safety-cases} discusses the benefits and downsides of using safety cases in industry self-regulation and government regulation of frontier AI. \Cref{components-of-a-safety-case} outlines what frontier AI safety cases could look like in more detail. \Cref{implementation-challenges} discusses potential challenges to using safety cases in frontier AI regulation. \Cref{policy-recommendations} provides recommendations for developers and policymakers. \Cref{conclusion} concludes with a summary of our main contributions and suggestions for further research.

\section{Safety cases and frontier AI}\label{safety-cases-and-frontier-ai}

This section outlines the concept of safety cases. We cover what safety cases are (\Cref{what-is-a-safety-case}), what safety cases for frontier AI might look like (\Cref{what-is-a-frontier-ai-safety-case}), and how they relate to existing frontier AI safety frameworks (\Cref{how-do-safety-cases-relate-to-existing-frontier-ai-safety-frameworks}).

\begin{table}[t!]
\small
\renewcommand{\arraystretch}{1.15}
\begin{tabular}{@{}p{0.4\linewidth}p{0.56\linewidth}@{}}
\toprule
\centering\bfseries Components &\centering\bfseries\arraybackslash Significance \\ \midrule
\renewcommand{\arraystretch}{1}\begin{tabular}[t]{@{}l@{}}
\textbf{Argument} \\ (“An argument…”)\end{tabular} & A safety case must comprehensively justify why a system is safe. It must explain the relevance and sufficiency of the available evidence. \\\\[-6pt]
\renewcommand{\arraystretch}{1}\begin{tabular}[t]{@{}l@{}}\textbf{Evidence} \\ (“…supported by evidence…”)\end{tabular} 
& A safety case must provide support for claims and clearly state its assumptions. It must document how safety has been assessed and achieved. \\\\[-6pt]
\renewcommand{\arraystretch}{1}\begin{tabular}[t]{@{}p{\linewidth}@{}}\textbf{Objectives} \\ (“…that a system is safe enough…”)\end{tabular} & A safety case must ultimately be about outcomes (“the system is safe enough”) rather than a product (e.g. “the system design adheres to safety standards”) or process (e.g. “the system was tested with SOTA techniques”).\\\\[-6pt]
\renewcommand{\arraystretch}{1}\begin{tabular}[t]{@{}l@{}}\textbf{Scope} \\ (“…in a given operational context.’’)\end{tabular} & A safety case must specify the conditions under which the argument is valid. \\ \bottomrule
\\[-4pt]
\end{tabular}
\caption{Summary of the four key components of a safety case}\label{tab:1}
\end{table}

\subsection{What is a safety case?}\label{what-is-a-safety-case}

A safety case is a structured argument, supported by evidence, that a system is safe enough in a given operational context \citep{stan1996}.\footnote{We use the term “operational context” in a broad sense, meaning roughly “setting in which the model is used”. In the context of frontier AI, key elements of the operational context include who can access the model and in what ways. Safety cases for frontier AI do not have to be restricted to a specific use case; in fact, the relevant operational context will likely often be a widespread deployment context in which many users can use the system in open-ended ways.} Safety cases are often used to inform major “go/no go” decisions, such as whether to build, deploy, procure, or license a system. They typically inform such decisions via a three-staged process, illustrated in \Cref{fig:1}. First, the safety case is produced by the developer. Then, it is reviewed by an internal or external actor. Finally, it is shared with key decision-makers. Sometimes, the same actor (e.g. a regulator) acts as both reviewer and decision-maker. In some contexts, a safety case is produced already before a system is developed and continually assessed and updated throughout its lifecycle.

Safety cases are a framework for assessing and communicating about the safety of a system. Relative to its alternatives, the defining feature of a safety case is that it is an \textit{argument }about what \textit{outcomes }have been achieved – namely, that the system does not pose unacceptable risk. A safety case is not just a collection of decision-relevant information, nor is it a checklist of best practices followed by the developer. Rather, a safety case explains why the information presented provides sufficient assurance that the system is safe enough \citep{kelly2017}.

Safety cases have become increasingly common in recent decades because they are seen as more comprehensive and flexible than traditional approaches to assurance. They emerged in sectors such as energy, petrochemicals, and transportation, where assurance traditionally meant adhering to specific product design rules \citep{leveson2011}. In response to accidents in the 1960s-80s, this rules-based approach was questioned on the grounds that it did not encourage developers to comprehensively assess the safety of their systems \citep{kelly2017}. Safety cases emerged as an alternative approach. Since then, safety cases have spread to a variety of industries, including defense \citep{2020}, aerospace \citep{Dezfuli2015-au},  and more recently AI-related industries such as security \citep{alexander2017}, software \citep{islam2020}, and autonomous vehicles \citep{favaro2023}. However, safety cases have also been criticized for providing a false sense of assurance when in reality it is very difficult to produce an adequate safety case or review it effectively \citep{langari2013,leveson2011,wassyng2011}. While there is little empirical evidence for the efficacy of safety cases \citep{habli2021}, and such evidence is difficult to produce given the nature of risk \citep{koopman2022}, practitioners mostly consider them effective \citep{rinehart}, and they are a recognised best practice in the UK \citep{cleland2012evidence}. 

A safety case has four key components: Objectives, arguments, evidence, and scope. It aims to show that certain safety objectives have been achieved (e.g. that risk is below a specific threshold). The safety case must present one or more arguments for why those objectives have been achieved, with the assumption that the system is unsafe unless convincing arguments are made to the contrary. These arguments must be supported by evidence (e.g. test, evaluation, validation, and verification results). Finally, the safety case must specify the scope in which it is valid (e.g. a specific deployment context with specific safety measures in place). \Cref{tab:1} outlines these four components and their significance for the unique assurance approach of safety cases.

\begin{figure}[t!]
    \centering
    \includegraphics[width=\linewidth]{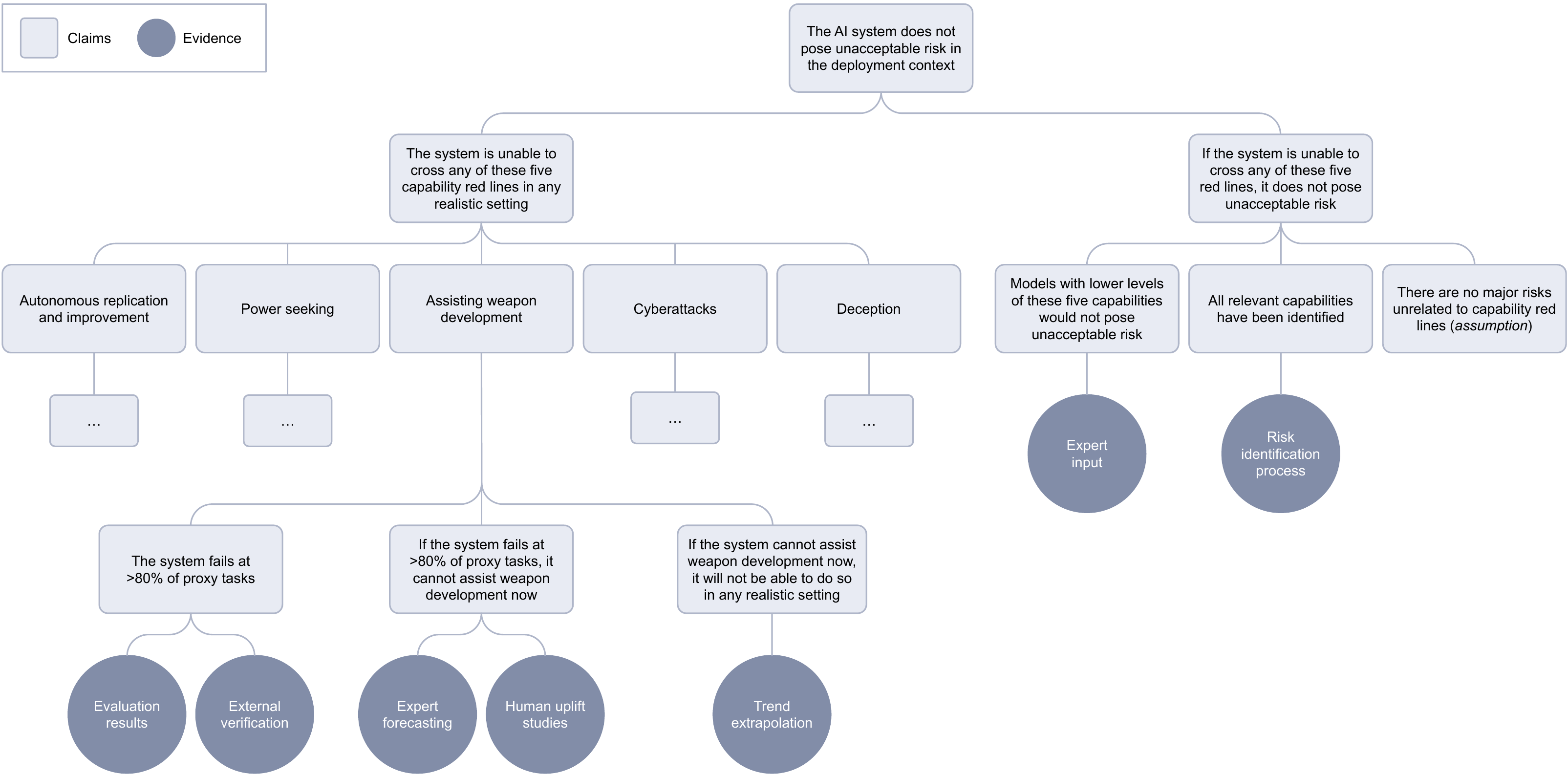}
    \vspace{-0.5em}
    \caption{Sketch of a safety case argument}
    \label{fig:2}
\end{figure}

\subsection{What is a frontier AI safety case?}\label{what-is-a-frontier-ai-safety-case}

To help give an intuitive sense of safety cases, we sketch what early-stage safety cases for a current frontier AI system may look like in \Cref{fig:2} and \Cref{tab:2}. As is typical for current frontier AI systems, the operational context is one of widespread deployment, where many users can interact with the system in an open-ended way. The sketch follows developers' existing safety frameworks in focusing on catastrophic risks enabled by certain dangerous capabilities. The sketch uses an ``inability argument'' \citep{clymer2024,goemans_forthcoming}. It claims that the system does not pose unacceptable risk because it is not capable enough to cause serious harm even if it attempts to. This claim is then broken down into a number of capability red lines \citep{Hinton2024} and providing evidence that the system does not cross these red lines. Note that this sketch is only meant to illustrate what a safety case might look like. We do not claim that a safety case with this structure or substance would be sufficient or sound, even for near-term systems.

\begin{table}[ht!]
\centering
\small
\renewcommand{\arraystretch}{1.2}
\begin{tabular}{@{}p{0.12\linewidth}p{0.28\linewidth}p{0.53\linewidth}@{}}\toprule
\centering\bfseries Components &
\centering\bfseries Explanation &
\centering\bfseries\arraybackslash Examples \\\midrule
{\bfseries Scope} &
  \textbf{System specification}: Information about the frontier AI system and deployment set-up (e.g. architecture, training process, safeguards, access). &
  \begin{minipage}[t]{\linewidth} 
  \begin{itemize}[leftmargin=*]
    \item Deep learning network with 1 trillion parameters. 
    \item Pre-training with $10^{26}$ total training FLOP using self-supervised learning on CommonCrawl web text.
    \item Fine-tuned with reinforcement learning from human feedback (RLHF) to be helpful, harmless, and honest.
    \item Model will be accessible via an API to anyone with a free online account.
    \item An AI assistant will monitor activity and flag suspected violations of our usage policy to a member of staff.
  \end{itemize}
  \end{minipage} \\\\[-2pt]
 &
  \textbf{Assumptions}: Conditions under which the safety case holds (e.g. temporal scope, assumptions about how the system will be used, assumptions about the effectiveness of safeguards). &
  \begin{minipage}[t]{\linewidth} 
  \begin{itemize}[leftmargin=*]
  \item We include a buffer of expected capability improvements from (a) one year of expected capability improvements from scaffolding and prompting, and (b) additional training using over $10\%$ of pre-training compute.
  \item Once these buffers have been surpassed, a new safety case must be prepared.
  \end{itemize}
  \end{minipage} \\\\[-2pt]
{\bfseries Objectives} &
  Safety requirements that the safety case will argue have been met. &
  \begin{minipage}[t]{\linewidth} 
  \begin{itemize}[leftmargin=*]
  \item The AI system does not pose unacceptable risk in the deployment context. Unacceptable risk is defined as a probability of  $\geq 10^{-7}/\,$year of causing an event with $\geq 1,000$ fatalities.
  \end{itemize}
  \end{minipage} \\\\[-2pt]
{\bfseries Arguments} &
  A hierarchy of claims that, if true, collectively imply that the objectives have been met; an explanation of the developer’s level of confidence that each claim is true and that the objectives have been met. &
  \begin{minipage}[t]{\linewidth} 
  \begin{itemize}[leftmargin=*]
  \item See~\Cref{fig:2}. 
  \end{itemize}
  \end{minipage} \\\\[-2pt]
{\bfseries Evidence} &
  Collection of all evidence sources that support the claims made in the argument; detailed explanation of how the evidence was produced, its results, and verification. &
  \begin{minipage}[t]{\linewidth} 
  \begin{itemize}[leftmargin=*]
  \item Risk identification report covering methodology, risk taxonomy, and external verification.
  \item Results of expert consultation to determine capability red lines covering methodology, names of experts, and results.
  \item Evaluation report covering methodology, results, and the evaluation script.
  \item Report produced by external red-teamers.
  \item Trend extrapolation analysis forecasting effect of post-deployment enhancements on capabilities.
  \end{itemize}
  \end{minipage}\\
  \bottomrule
  \\[-4pt]
\end{tabular}
\caption{Sketch of each safety case component}\label{tab:2}
\end{table}

\subsection{How do safety cases relate to existing frontier AI
safety
frameworks?}\label{how-do-safety-cases-relate-to-existing-frontier-ai-safety-frameworks}

Safety frameworks have been central in the conversation about frontier AI risk management so far. A safety framework is a developer's plan for assessing and mitigating risks posed by its AI systems \citep{dsit2024}. It is an organization-level policy that applies to all frontier systems, in contrast to a safety case which only applies to an individual system. 16 companies have committed to publishing safety frameworks ahead of the 2025 AI Action Summit \citep{dsit2024} and four companies have already done so \citep{anthropic2024,OpenAI2023-tt,googledeepmind2024,magic2024}. Existing safety frameworks focus on identifying dangerous capabilities, setting capability thresholds above which additional mitigations would be required, explaining how capabilities will be measured, and outlining mitigation options \citep{alaga2024,metr2024}.

A safety framework could form the basis of a safety case. For example, the central argument of a safety case could be that a system does not cross any of the capability thresholds identified in the developer's safety framework (similarly to \Cref{fig:2}), or that the developer has implemented the risk mitigations promised in the safety framework. However, to constitute a safety case, the argument must go beyond simply verifying that the developer has adhered to its framework.
The safety case must also justify the framework, explaining why adhering to it implies avoiding unacceptable risk. For example, the safety case must justify the choice of capability thresholds and argue that systems below the thresholds are safe enough to deploy in the chosen way.

Safety cases can play three roles in relation to safety frameworks: (1) Safety cases can be used to document that a developer has adhered to its safety framework with respect to a specific system. Such documentation is an important part of a safety framework. (2) Safety cases can allow safety frameworks to build in more flexibility. They are suitable for supporting subjective outcome-based claims such as “the safeguards are sufficient to prevent unacceptable risk”. They thus allow safety frameworks to use more outcome-based claims in cases where it is difficult to specify adequate pre-commitments. For example, highly specific pre-commitments may be inappropriate for more capable future systems, given that such systems are poorly understood and safeguards are rapidly evolving (3) Safety cases can help make safety frameworks more justified. They require developers to make implicit assumptions in safety frameworks explicit (e.g. that a system below certain capability thresholds does not pose unacceptable risk). While a good safety framework should contain such justification \citep{alaga2024}, safety cases provide a helpful prompt. Safety cases can thus both draw on prior work on safety frameworks and play a valuable role in their implementation.

\section{The case for safety cases}\label{the-case-for-safety-cases}

This section outlines the benefits and downsides of using safety cases in frontier AI governance. We consider two contexts in which safety cases could be used: industry self-regulation (\Cref{safety-cases-in-self-regulation}) and government regulation (\Cref{safety-cases-in-regulation}).\footnote{Safety cases could also be used in other contexts such as procurement. For example, safety cases are required by the UK Ministry of Defence for all equipment acquisitions \citep{stan1996}.} We conclude that, in both contexts, safety cases could be a useful and worthwhile tool for the highest-risk systems.

\subsection{Safety cases in self-regulation}\label{safety-cases-in-self-regulation}

\begin{figure}
    \centering
    \includegraphics[width=\linewidth]{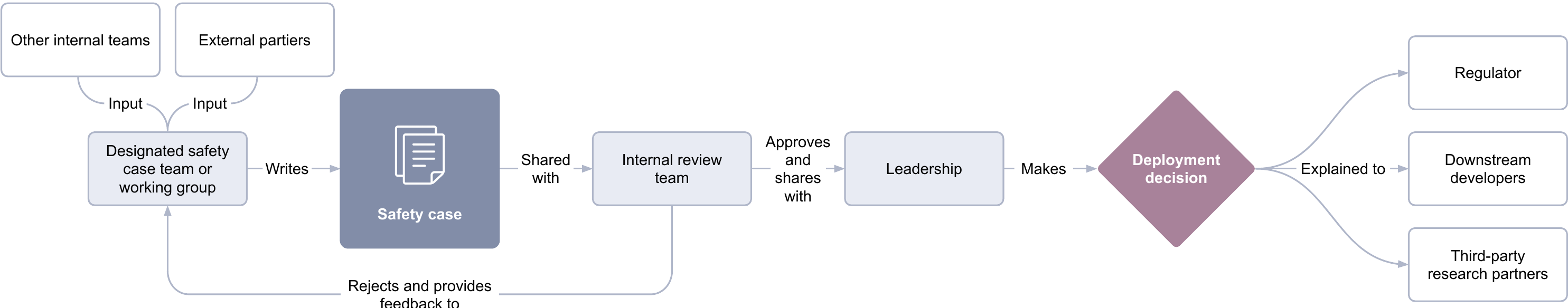}
    \vspace{-0.5em}
    \caption{Sketch of an internal safety case review process}
    \label{fig:3}
\end{figure}

If developers use safety cases internally as part of self-regulation,\footnote{By ``self-regulation'', we mean companies' internal policies, voluntary commitments or agreements, and industry-wide guidelines and processes (see \citep{baldwin2011,  coglianese2010}.} they could serve three purposes. First, safety cases could inform major decisions such as starting a training run or deploying a system. Senior management needs to assess if such decisions pose unacceptable risk; safety cases could be a key input into this assessment. They could be produced by a designated safety case team, evaluated and approved by an internal review team, and then shared with leadership who make the decision. A potential internal review process is outlined in \Cref{fig:3}. Second, safety cases could be used in ongoing risk management. Throughout the lifecycle, the development team can use a safety case as a framework for assessing how safe the system is and what additional safeguards are needed. Third, safety cases can be used to build trust with downstream developers, users, and governments, by providing assurance that the system is safe enough. Many companies in other industries, such as aerospace \citep{morris2001} and autonomous vehicles \citep{favaro2023}, use safety cases for all three purposes. They
contrast primarily against a less structured approach of sharing individual pieces of evidence or information.

One benefit of using safety cases internally is that they are an effective and scalable way to assess if a system is safe enough. Safety cases provide a way to integrate different sources of evidence into a single, comprehensive assessment of a system's safety \citep{cleland2012evidence,rinehart}. Moreover, safety cases are flexible enough to be applicable to a wide range of systems. They do not require specific risk assessment or mitigation techniques, but rather invite the developer to use whatever methods are most suitable to a particular system. This is particularly beneficial in the frontier AI context, as future systems may require very different types of assurance arguments to current systems (see \Cref{components-of-a-safety-case}).

Another benefit of using safety cases internally is that they are a useful tool for thinking and communication. Safety cases make otherwise implicit arguments explicit, which has numerous benefits. First, it can highlight reasoning flaws or assurance gaps \citep{cleland2012evidence,sujan2016}. This is particularly important in the frontier AI context because risk assessment is still novel and therefore liable to contain errors \citep{schuett2024-ia}. Second, it makes it easier to see how overall system safety is affected if the evidence changes (e.g. if a safeguard stops being effective) \citep{inge2007}. This is particularly important in the frontier AI context because new capabilities, use cases, or risks may emerge after deployment \citep{obrien2024}. Third, it can help stakeholders communicate more clearly about disagreements (e.g. about the validity of an argument or the strength of evidence) by creating a shared framework and set of assumptions \citep{cleland2012evidence,rinehart}. Fourth, it can aid communication by explaining the context and relevance of the evidence for a system’s safety, facilitating an understanding of both \emph{if} and \emph{when} the system is safe. This can be critical for decision-makers who are not intimately familiar with the system.

One downside of using safety cases internally is that they add costs of two kinds. First, safety cases may add risk management costs, in that they presuppose a reasonably high standard of risk management. However, these costs only apply to developers who do not already meet such a high standard. For example, safety cases should not add significant risk management costs for the companies that already meet their commitment to effectively and transparently manage risks posed by their systems \citep{dsit2024}. Second, safety cases add documentation costs in terms of keeping records of relevant information and structuring this information into an explicit argument. These costs can be reduced with measures such as safety case templates \citep{bloomfield2021} and cross-industry guidance and standards \citep{ulstandards&engagementulse2023}.

\begin{table}[p]
\small
\renewcommand{\arraystretch}{1.2}
\renewcommand{\tabcolsep}{3.5pt}
\begin{tabular}{>{\bfseries}p{0.15\linewidth}>{\bfseries}p{0.15\linewidth}p{0.3\linewidth}p{0.3\linewidth}}\toprule
&& \centering\bfseries Specific rules & \centering\bfseries\arraybackslash Safety cases \\\midrule
\multirow{3}{*}{\parbox{\linewidth}{\vspace{-1.85mm}\setlength{\baselineskip}{1.04\baselineskip}\raggedright Benefits of safety cases relative to rules}} &
  Flexibility and durability & It may be hard for regulators to specify a set of adequate safety practices. Rules might get outdated  and be hard to update, locking in suboptimal safety practices. & Regulators can require safety cases without having a clear sense of what specific safety practices are adequate for current or future systems. \\
 &
  Innovation & Developers lack incentives to develop safety practices beyond what is codified. & Developers are incentivised to develop and use more effective and efficient safety practices as this will allow them to produce better safety cases. However, this requires that regulators are capable of reviewing novel practices. \\
 &
  Allocation of responsibility & Regulators are responsible for identifying adequate safety practices, which may be difficult given they have more limited information, resources, and AI expertise. & Developers are responsible for specifying adequate safety practices, which is efficient given that they tend to have more information, resources, and AI expertise. \\\midrule
\multirow{3}{*}{\parbox{\linewidth}{\vspace{-1.85mm}\setlength{\baselineskip}{1.05\baselineskip}\raggedright Downsides of safety cases relative to rules}} &
  Transparency & Rules are typically public, clear-cut, and manageable in length, making them easier for third parties to scrutinize. & Safety cases may not be fully public, and may be long and technical. This may make it difficult for third parties to scrutinize whether regulators are reviewing safety cases appropriately. However, regulators can facilitate scrutiny by publishing guidance on how they review safety cases, explaining individual decisions, or subjecting themselves to auditing. \\
 &
  \mbox{Consistency of} enforcement & Compliance is assessed in a consistent, objective way. & Regulator judgment plays a key role in assessing compliance.\\
 &
  \mbox{Cost for} \mbox{developer} & It is often less costly for developers to demonstrate compliance with rules, and rules provide more legal certainty. However, if rules are poorly specified, they could lead to higher compliance costs by requiring inefficient or unnecessary safety practices. & It is resource-intensive for developers to produce safety cases. Because safety case review is subjective, safety cases also introduce more legal uncertainty for developers. \\\midrule
\multirow{3}{*}{\parbox{\linewidth}{\vspace{0.79mm}\setlength{\baselineskip}{1.05\baselineskip}\raggedright Factors that could be either a benefit or downside of safety cases relative to rules}} &
  Gamability & Rules might create loopholes or gaps. & Because safety case review is subjective, inadequate safety cases might sometimes pass review. \\

 &
  Risk of regulatory capture & A rulemaking process may be an easier target for industry lobbying relative to a safety case review process.& The subjective judgements required by safety case review might leave more room to give undue favors.
Regulators may also rely more on industry for information about safety practices and develop less independent expertise when they do not need to specify adequate safety practices. 
\\&
  \mbox{Cost for} \mbox{regulator} & It is easier to verify compliance, as it is typically relatively clear-cut if a rule has been followed or not. However, it may be more resource-intensive to develop the rules in the first place and to ensure they are up-to-date. & It is resource-intensive for regulators to verify compliance as it requires a bespoke and subjective review. However, set-up costs might be lower as it might be easier to describe an adequate safety case than to set specific rules.\\\bottomrule
  \\[-4pt]
\end{tabular}
\caption{Benefits and downsides of safety cases relative to specific rules}\label{tab:3}
\end{table}

\subsection{Safety cases in
regulation}\label{safety-cases-in-regulation}

Safety cases could also play several roles in frontier AI regulation\footnote{By ``regulation'', we mean laws, rules, and guidelines set by governments; industry self-regulation is not included. For more information on frontier AI regulation, see \citeA{Anderljung2023} and \shortciteA{schuett2024}.} First, safety cases could support information sharing and transparency \citep<see>{kolt2024}. Regulators could require require that developers submit safety cases when deploying frontier models, without regulators having any formal powers to penalize or restrict deployment. This could be considered an extension of existing reporting requirements \citep{The_White_House2023-ru}. Such information sharing could help regulators build capacity for effectively governing AI more broadly. For example, they could help regulators write rules and standards for AI systems behind the frontier by keeping them informed about state-of-the-art safety practices. An information sharing requirement could also be a flexible interim measure, allowing regulators to build experience with safety case review while only scaling up enforcement powers if there is evidence that future frontier models pose serious risk. 

Second, safety cases could be used to assess compliance with regulation requiring frontier AI developers to manage risks posed by their systems. Such regulation is already in place in the EU \citep{euparliament2024} and may also be adopted in other jurisdictions. Regulators could take two broad approaches to assessing compliance with such regulation. First, they could set specific rules about how developers should assess and mitigate risk (e.g. specific safety tests or training techniques that must be used). Alternatively, they could allow developers to choose how to assess and mitigate risk, and use safety cases to assess if developers have done enough.\footnote{Regulators could also assess compliance by requiring a variety of information from developers without requiring that it be organized as a safety case. The relative benefits and downsides of using safety cases are similar to those outlined in \Cref{safety-cases-in-self-regulation}.} This is a common use of safety cases in industries such as nuclear power  \citep{ukofficefornuclearregulationonr2021,usnuclearregulatorycommissionnrc2007}, on-shore and off-shore petrochemical installations \citep{ukparliament2015,ukparliament2015a}, and rail operators \citep{draganjovicic2009,ukparliament2000}. In these industries, a regulator must approve a safety case before granting a license to build or operate the relevant facilities. However, safety cases could also be combined with other enforcement mechanisms such as liability. For example, regulators could use safety cases to retrospectively assess if a developer was negligent in deploying a certain model.

Using safety cases to assess compliance with regulation has both advantages and disadvantages relative to using specific rules. These are summarized in \Cref{tab:3}. There is no clear-cut answer to which approach is preferable in the abstract. Rather, this depends on the specific characteristics of the regulator and the industry being regulated. \Cref{tab:4} outlines some of the relevant characteristics and discusses the extent to which they hold true in the context of frontier AI regulation. Both tables are based on the literature on safety cases \citep{cleland2012evidence,hawkins2013,inge2007,rinehart,sujan2016}, and principles-based regulation more broadly \citep{coglianese2020,Decker2018-kd,schuett2024}.

\begin{table}[t!]
\renewcommand{\arraystretch}{1.15}
\setlength{\tabcolsep}{6pt}
\small
\begin{tabular}{p{0.3\linewidth}p{0.3\linewidth}p{0.3\linewidth}}\toprule
\bfseries Conditions\footnotemark &
\bfseries Why are safety cases preferable in this condition? &
\bfseries How likely is it that this condition holds for frontier AI? \\\midrule
\textbf{Difficult to specify rules:} Experts have little confidence in what precise actions will reliably reduce risk to an acceptable level. &
  Rules are more likely to be lacking and could therefore lock in insufficient or unnecessary safety practices. If there are not well-established best practices, it is also especially important to incentivise innovation in safety practices. &
  \textbf{Likely:} Frontier AI systems are novel and rapidly evolving; highly general-purpose; complex and not mechanistically understood; heterogeneous and lacking in standard designs. These factors make it hard to specify rules. \\\\[-4pt]
\textbf{Rule lock-in:} Once rules have been established, changing them is a difficult and lengthy process. &
  Rules are more likely to become outdated and lock in suboptimal safety practices. &
  \textbf{Unclear:} Depends on jurisdiction. \\\\[-4pt]
\textbf{Information asymmetry:} Developers have significantly better access to relevant information and expertise compared to regulators. &
  It is more efficient to place the burden on developers to identify adequate safety practices. &
  \textbf{Likely:} AI expertise is naturally concentrated in frontier AI companies, though several governments have established AI Safety Institutes to boost their capacity. Developers naturally have more and earlier information about frontier AI, and some tacit knowledge may be difficult to convey even with information sharing requirements. \\\\[-4pt]
\textbf{High risk:} The regulated activity poses high risk, defined in terms of the likelihood and impact of the product causing harm. & It is more likely worth imposing a higher compliance cost on developers to reduce risk. &
  \textbf{Possible:} There is significant expert disagreement about the risks posed by frontier AI. However, it seems difficult to rule out that future AI systems may pose high risk. \\\\[-4pt]
\textbf{Few covered developers:} There are few developers and systems that are in scope for the safety case requirement. & The higher compliance cost is limited to a few developers; the enforcement cost for the regulator is lower; and safety case review is likely to be more prompt and thorough. &
  \textbf{Likely:} Frontier AI development is currently resource-intensive in terms of compute, energy, capital, and specialized expertise. There are therefore only a handful of companies developing frontier AI. However, it is unclear if this will continue to be true. \\\\[-4pt]
\textbf{Covered developers can absorb costs:} Covered developers are more well-resourced and able to bear the compliance costs and legal uncertainty of safety cases. &
  It is more likely worth imposing a higher compliance cost on developers and less likely that compliance costs will stifle innovation. &
  \textbf{Likely:} Because frontier AI development is resource-intensive, only relatively well-resourced companies are likely to be covered.\\\bottomrule
  \\[-4pt]
\end{tabular}
\caption{Conditions under which safety cases are preferable to
precise rules}\label{tab:4}
\end{table}
\footnotetext{Some conditions that seem relevant, but do not clearly favor either safety cases or specific rules, include: (1) Regulator capacity: Rules seem to require more regulator capacity at the point of writing the regulations, and principles at the point of enforcing the regulations; (2) Degree of alignment between regulator and regulated objectives: If the regulated entities are incentivised to try to circumvent the regulatory objective, it is unclear if rules (which may have gaps and loopholes) or safety cases (which are subjectively assessed) are easier to ``game''.}

We conclude that the conditions of the frontier AI industry favor the use of safety cases to assess when deployment poses unacceptable risk. First, given that the frontier AI industry is so rapidly evolving, complex, and poorly understood, a more flexible approach to assessing safety seems suitable. Second, safety cases could capitalize on the expertise and resources of frontier AI developers. Third, given that frontier AI systems are currently produced by only a handful of relatively well-resourced developers, the compliance costs seem bearable. However, regulators may want to complement safety cases with specific rules in some cases. For example, regulators could specify rules in better-understood risk domains or mandate some particular safeguards. Regulators could also continuously replace safety cases with precise rules as the frontier moves, using safety cases to identify best practices that are then applied to future systems at a similar capability level. Finally, regulators should consider the implementation challenges outlined in \Cref{implementation-challenges} when deciding whether to use safety cases.

\section{Components of a safety case}\label{components-of-a-safety-case}

This section outlines the content of frontier AI safety cases. We cover the four key components of a safety case: scope (\Cref{scope}), objectives (\Cref{objectives}), arguments (\Cref{arguments}), and evidence (\Cref{evidence}). For each of these components, we discuss what is feasible today and what developments could happen in the future. We conclude that developers can already produce safety cases with existing techniques, but that research breakthroughs will likely be needed to produce compelling safety cases for more capable future systems.

\begin{figure}[t!]
\begin{tcolorbox}[boxrule=1pt,enhanced jigsaw, sharp corners,pad at break*=1mm,colbacktitle=boxbackground,colback=boxbackground,colframe=boxline,coltitle=black,toptitle=1mm,bottomtitle=1mm,width=\linewidth,fonttitle=\bfseries,parbox=false,titlerule=1pt,title=Box 1: Safety cases throughout the system lifecycle
,phantom={\phantomsection\hypertarget{box1}}]

According to best practice, a safety case is not just a one-off event, but a living document to be developed and maintained across a system's lifecycle \citep{kelly2017,koopman2022}. This is illustrated in \Cref{fig:4}.

\begin{minipage}{\linewidth}
\vspace{1.5\baselineskip}
    \centering    \includegraphics[width=\linewidth]{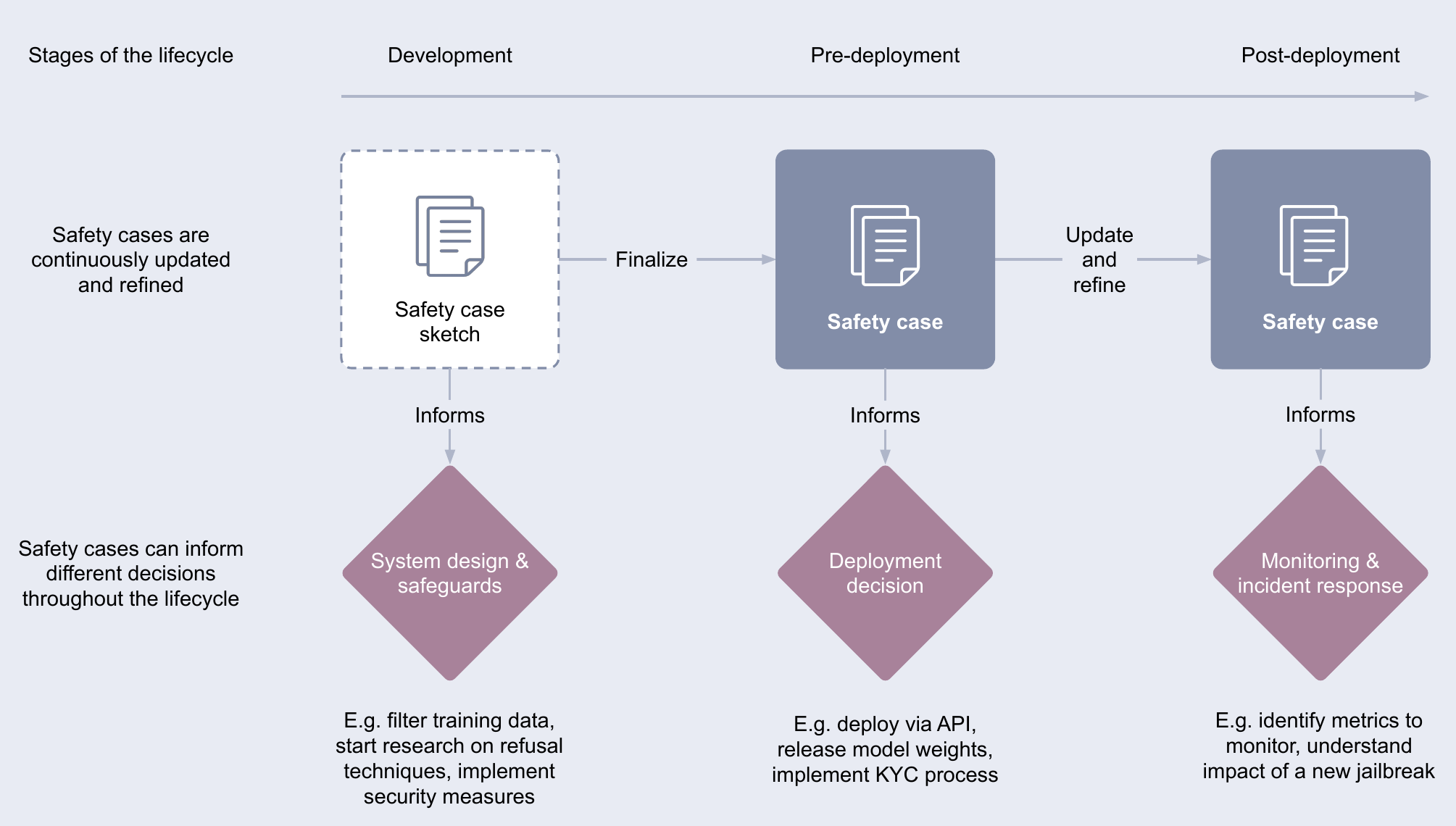}
    \captionof{figure}{The stage and role of safety cases at each stage of the system lifecycle}
    \label{fig:4}
\vspace{1.5\baselineskip}
\end{minipage}

During development (e.g. system design, pre-training, and fine-tuning), the developer should gradually build the safety case. The goal is to ensure that safety is embedded in the design of the system rather than tagged on at the end, and to make it easier to produce the pre-deployment safety case. The developer should sketch the intended safety case argument and use it to identify model-level safeguards (e.g. via data filtering or reinforcement learning). The sketch can also help identify potential gaps where more work is needed to assure the safety of the system. Finally, the developer should keep records of evidence that may be used in the safety case. In industries where safety cases inform licensing decisions, such as nuclear energy \citep{ukofficefornuclearregulationonr2021}, it is common for the regulator to be involved throughout the process, providing guidance and feedback on early drafts.\\

After deployment, the developer should continue to update the safety case. New risks may emerge as the system is modified or used in unexpected ways, and real-world information may invalidate the developer's risk assessment. Developers should therefore track key metrics or risk indicators that provide evidence for claims in the safety case \citep{koopman2022}. For example, if the safety case estimates that a harm refusal technique has a certain success rate, the developer should track the actual success rate after deployment. Developers should also identify conditions that would require an update to the safety case. Updates could happen at regular intervals and/or in response to triggers such as a new jailbreak or significant post- eployment
enhancements.
\end{tcolorbox}
\end{figure}

\subsection{Scope}\label{scope}

The scope is the range of conditions under which the safety case holds.\footnote{Our definition of “scope” is based on the definition of “context” in \citeA{kelly2017}.} The scope should include: (1) A detailed specification of the system (e.g. its architecture, training process, and safeguards) and intended deployment context (e.g. whether the model weights will be released or the model will be accessible only via API). (2) An overview of which changes to the system (e.g. post-deployment fine-tuning) or deployment context (e.g. expanding access) are considered within the scope of the safety case and which would require an updated safety case. (3) Any other scope restrictions, in particular the temporal scope (e.g. the safety case being valid for two years), assumptions (e.g. a safeguard being resistant to jailbreaking), or out-of-scope use cases (e.g. use in medical diagnosis).\footnote{The safety case may need to address how the developer will prevent the system from being used for out-of-scope purposes.} The scope is relevant for reviewers who should assess if the safety case rests on reasonable assumptions. It may also be relevant to audiences such as governments or downstream developers in avoiding unsafe uses or addressing risks not covered by the safety case. 

A particular challenge of frontier AI safety cases is that their scope will tend to be broad. Frontier AI systems are highly general-purpose and have so far been deployed widely for open-ended use. This makes it challenging to assure safety, as it is virtually impossible to assess all possible risk scenarios individually \citep{anwar2024}.  As frontier AI systems become more capable, assurance may become even more difficult (see \Cref{arguments}). As such, safety cases may need to be restricted to a narrower scope (e.g. producing cases for specific applications rather than general release). On the other hand, risk assessment and mitigations may advance enough that it is possible to maintain a broad scope. Insofar as assurance is possible, it is both in the commercial interest of developers and the interest of society for developers to widen the scope of safety cases so that systems can be used for a wider range of beneficial applications.

\subsection{Objectives}\label{objectives}

Objectives are requirements that operationalise what it means for the system to be safe enough to deploy \citep{kelly2017}. A typical objective is a risk threshold, such as a specified probability of a given level of harm (e.g. a probability of $\geq 10^{-7}/\,$year of causing an event with $\geq 1,000$ fatalities). It is common in other industries to set multiple risk thresholds: One region for ``unacceptable risk'' which must not be exceeded, one region in which risks must be ``as low as reasonably practicable'', i.e. where all reasonable mitigations must be implemented, and one region for ``acceptable'' risk which does not require further mitigations \citep{koessler2024}. Objectives can also be comparative. For example, a common objective in autonomous vehicles safety cases is that a self-driving car must be at least as safe as a human driver \citep{koopman2022}. In a regulatory context, objectives are typically set by regulators, but can also be selected by developers. In the latter case, reviewers should assess the choices of objectives.

Over time, the objectives of frontier AI safety cases could become more specific, quantitative, and directly based on risks. Early safety cases may initially use broad, qualitative objectives (e.g. that the system does not pose unacceptable risk in the deployment context).
Objectives could also be restricted to only a few key risk domains, rather than global risk. Finally, early safety cases could use proxy-based objectives, which do not measure risk directly but measure other outcomes indirectly related to risk, such as capability thresholds \citep{koessler2024}. As safety case methodology develops, objectives should likely move towards direct measurements of risk. They could also increasingly contain sub-objectives for specific risks as risk analysis improves, as is done in nuclear energy \citep{nrc_safety_goals_1986}.

\subsection{Arguments}\label{arguments}

\begin{figure}
    \centering    \includegraphics[width=0.65\linewidth]{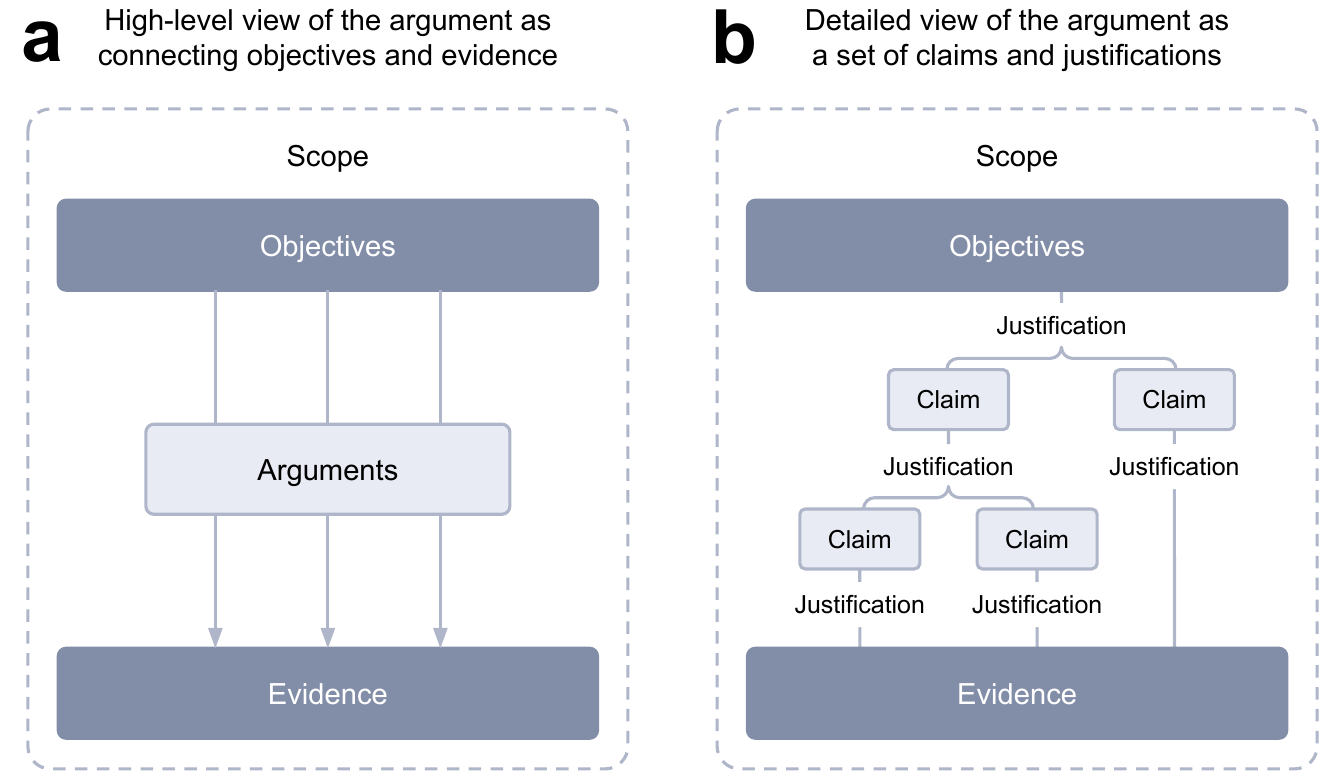}
    \vspace{0.25em}
    \caption{Definition of a safety case argument}
    \label{fig:5}
\end{figure}

Arguments explain why the evidence gives sufficient reasons to think that the objectives have been met \citep{kelly2017}.\footnote{The term ``argument' is used in a number of different ways in the safety case literature. For example, in the Claims Arguments Evidence framework, it refers to the justification linking a specific claim to a specific piece of evidence \citep{adelard2024}; and in propositional logic, an argument would include the evidence and conclusion (i.e the claim that the objective has been met). We follow the use of the term in the Goal Structuring Notation framework \citep{scscassurancecaseworkinggroupacwg2021}.} Arguments break the objective down into subclaims that, if true, collectively imply that the
objective has been met, and they explain what evidence supports the subclaims. This is illustrated in \Cref{fig:5}. For a concrete example, see \Cref{fig:2}.
Reviewers must assess if the argument is valid (i.e. the subclaims imply that the objective has been met) and if the subclaims are adequately supported by the evidence.\footnote{Claims are rarely ever proven or demonstrated with certainty; rather, evidence provides some reason to think the claim is true, but these reasons may be overridden by defeaters \citep{bloomfield2021}. Some approaches to safety cases model uncertainty more explicitly and systematically \citep{bloomfield2010,denney2011,nesic2021,wang2018}.}

\begin{figure}
    \centering
    \includegraphics[width=\linewidth]{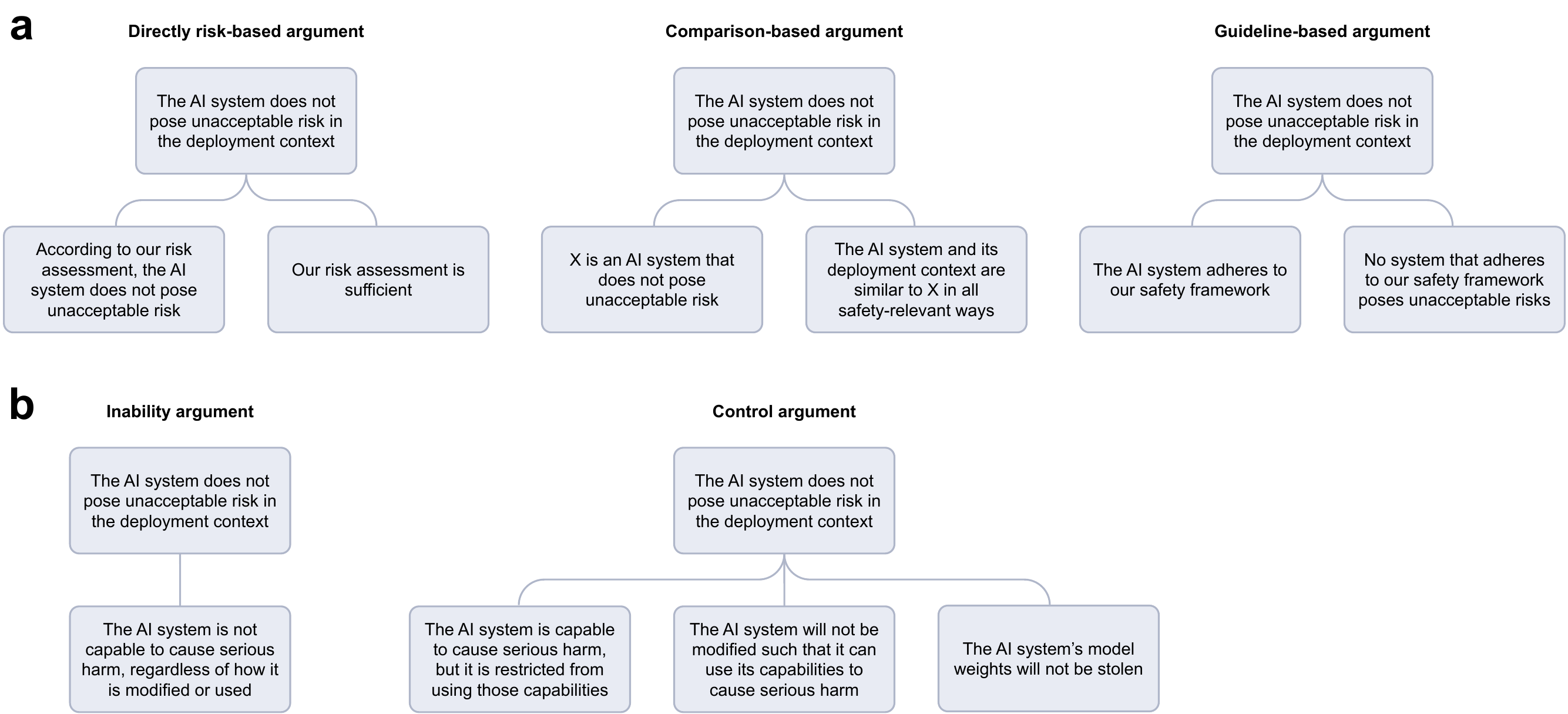}
    \vspace{-0.2em}
    \caption{Different types of arguments for frontier AI safety cases}
    \label{fig:6}
\end{figure}

There can be many ways to argue for the same objective and developers can choose which argumentative strategy to use. For example, safety case arguments could be based on direct risk estimates, comparisons with other systems, or guidelines, as sketched in \Cref{fig:6}a.\footnote{These three categories are inspired by the three risk acceptance principles of the European Rail Agency, i.e. the allowed three methods for demonstrating that the risk of a given railway system is acceptable \citep{draganjovicic2009}.} Within explicit risk estimates, developers can also adopt multiple strategies. For example, developers can argue that a given system is not capable enough to cause serious harm (inability), that control measures prevent it from causing serious harm (control), that it reliably does not cause serious harm (trustworthiness), or that other AI systems have verified that it will not cause serious harm (deference) \citep{clymer2024}.\footnote{Capabilities are far from the only risk factor, and safety cases may also need arguments not closely tied to capabilities, for example about the risk of system failure in high-risk contexts or the diffuse effects of widespread use of the system. However, we focus primarily on capability-related arguments, since those are the focus of developers’ existing safety frameworks \citep{anthropic2024,OpenAI2023-tt,googledeepmind2024,magic2024}}
\Cref{fig:6}b contrasts an inability and a control argument. While developers choose the arguments, regulators may publish guidance on standard arguments they deem acceptable or preferable.

Early frontier AI safety cases will likely rely on inability arguments. An example inability argument, based on existing safety frameworks (as described in \Cref{what-is-a-frontier-ai-safety-case}), could claim that a system is safe because it does not cross any of a number of capability thresholds. The meat of the argument then consists in justifying the capability thresholds and motivating the internal and external validity of capability evaluations. Inability arguments are relatively well-established. Identifying and evaluating dangerous capabilities is a major focus of existing safety frameworks \citep{anthropic2024,OpenAI2023-tt,googledeepmind2024,magic2024}, system cards arguably already make inability arguments \citep{anthropic2023,openai2024} and researchers are developing early-stage sketches of their explicit structure \citep{goemans_forthcoming}. There are still open questions, such as how to incorporate post-deployment enhancements \citep{davidson2023}, account for defensive uses of capabilities \citep{mirsky2023}, or address risks that are less closely tied to dangerous capabilities such as systemic risks \citep{zwetsloot2019} or risks from AI malfunction \citep{raji2022} that are less closely tied to dangerous capabilities. Nonetheless, it is possible for developers to make reasonable inability arguments already now, and they may be a useful starting point for developing safety case methodology.

In the long run, however, inability arguments are unlikely to be the main focus of safety cases. If frontier AI capabilities continue to accelerate, systems will eventually cross dangerous capability thresholds, and so safety cases must instead argue that those capabilities will not be used to cause unacceptable harm. Other arguments, such as control, trustworthiness, and deference arguments, cannot yet be supported via well-established methods, as described in \Cref{evidence}, and have not yet been stress tested in the real world. By the time safety cases rely on these arguments, the stakes will also be higher, assuming that systems with more dangerous capabilities pose greater risk. For these reasons, developing alternative arguments is a key condition for safety cases to be useful in the long run. Some research on alternative argument structures is already underway \citep{irving2024}.

\subsection{Evidence}\label{evidence}

Evidence consists of observable facts that support the claims made in the arguments \citep{kelly2017}. Frontier AI safety cases will likely use a wide variety of evidence types. Some examples include \emph{empirical tests} (e.g. dangerous capability evaluations [\citeNP{phuong2024,shevlane2023}], deployment experiments [\citeNP{weidinger2023}], or simulation exercises to stress test preparedness plans), \emph{mathematical models} (e.g. predictions of post-deployment capability gains [\citeNP{ganguli2022b}]), \emph{formal verifications or proofs} that the system satisfies certain requirements \citep{dalrymple2024}; \emph{expert judgements} (e.g. results from surveys or interviews), and \emph{documentation of plans, policies, and processes} (e.g.  internal governance structures [\citeNP{schuett2023,schuett2024-ia}] or incident response plans [\citeNP{obrien2024}]). Reviewers should both evaluate the relevance and strength of the evidence (i.e. to what extent it supports the claims) and selectively verify individual pieces of evidence (e.g. reproducing select capability evaluations).

There are already a number of techniques available to evidence inability arguments, though some of them are flawed and could be improved with further research.  A core challenge of inability arguments is to justify the choice of capabilities and capability thresholds. Current evidence techniques include expert elicitation \citep{murrayetal.}, threat modeling, public input \citep{thecollectiveintelligenceproject2023}, human uplift studies \citep{weidinger2023}, and forecasting \citep{phuong2024}. These techniques are relatively early-stage. Over time, safety cases may use more detailed, robust, and externally verified versions of these techniques. For example, theoretical threat models may increasingly be replaced by quantitative estimates backed by empirical evidence. Another core challenge of inability arguments is to assess model capabilities. Developers and third parties have already developed a number of evaluations \citep{kinniment2024,phuong2024}, and system releases tend to be accompanied by internal and external evaluations \citep{anthropic2023,openai2024}. Again, evaluations are early-stage and there is significant disagreement about their validity \citep{anwar2024,weidinger2023,rauh2024}. Nonetheless, it is plausible that these techniques are adequate for current systems, and further iteration could make them more robust.

There is much less scientific understanding of how to evidence arguments beyond inability arguments. It could become challenging to rely on empirical evidence about system behavior as systems may strategically modify their behavior to pass evaluations \citep{hubinger2024,ngo2022,weij2024}. Safety cases may therefore increasingly need to rely on a mechanistic understanding of systems, for example based on evidence from model internals or the training process \citep{wasil2024}. Even developers do not currently have such a mechanistic understanding \citep{anwar2024}. Various research directions may help provide evidence sources for other kinds of safety case arguments. Some examples include mechanistic interpretability \citep{bereska2024}, formal verification \citep{dalrymple2024}, control \citep{greenblatt2024}, and automated AI safety \citep{bai2022,burns2024,christiano2018,irving2018,leike2018}. However, these fields are nascent, it is still unclear if they will produce robust evidence techniques, and doing so will likely require significantly more investment.

\section{Implementation challenges}\label{implementation-challenges}

This section discusses potential challenges to using safety cases in frontier AI governance. We discuss two technical challenges, i.e. challenges relating to the content of safety cases (\Cref{technical-challenges}), and three institutional challenges, i.e. challenges relating to the organizations and processes by which safety cases are reviewed (\Cref{institutional-challenges}). In each section, we explain how the challenges might affect the use of safety cases in both self-regulation and regulation. Table 5 summarizes the key technical and institutional challenges. If these challenges are left unaddressed, the use of safety cases may be inadvisable. It could result in the conclusions of safety cases being unreliable, biased, and overly relied upon.

\begin{table}[t!]
\small
\renewcommand{\tabcolsep}{7.5pt}
\centering
\begin{tabular}{p{0.14\linewidth}p{0.365\linewidth}p{0.385\linewidth}}\toprule
 &
  \centering\bfseries Technical challenges &
  \centering\bfseries\arraybackslash Institutional challenges \\\midrule
\textbf{Self-regulation} &
 \begin{minipage}[t]{\linewidth}
  \begin{itemize}[leftmargin=*]
    \item Building consensus on safety case methodology
    \item Developing safety cases for more capable future systems
  \end{itemize} 
  \end{minipage}&
  \begin{minipage}[t]{\linewidth}
  \begin{itemize}[leftmargin=*]
    \item Implementing an appropriate internal review process
    \item Incorporating third parties into safety case review
  \end{itemize}
  \end{minipage}\\\\
\textbf{Regulation} &
  \begin{minipage}[t]{\linewidth}
  \begin{itemize}[leftmargin=*]
    \item Setting an appropriate bar for what constitutes an adequate safety case
  \end{itemize}
  \end{minipage}&
  \begin{minipage}[t]{\linewidth}
  \begin{itemize}[leftmargin=*]
    \item Appointing or establishing a body to receive safety cases securely 
    \item Building capacity and expertise to effectively review safety cases
    \item Incorporating third parties into safety case review
  \end{itemize}
  \end{minipage}\\\bottomrule
  \\[0pt]
\end{tabular}
\caption{Overview of key technical and institutional challenges to using safety cases to inform decision-making}
\end{table}

\subsection{Technical challenges}\label{technical-challenges}

The first technical challenge is to develop and build consensus on methodology for frontier AI safety cases. This is a particular challenge for three reasons: (1) The idea of applying safety cases to frontier AI is novel and there is not yet any well-established methodology. (2) 
Frontier AI has many characteristics of complex systems\footnote{Complex systems are systems that are fundamentally difficult to understand (e.g. due to the interactions of their parts meaning that components cannot be analyzed separately). Complexity has been defined as intellectually unmanagability \citep{leveson2012} or generating outputs with high statistical complexity \citep{ladyman2013}.}, which makes assurance particularly challenging. For example, frontier AI is opaque (i.e. not understood at a mechanistic level) and general-purpose (i.e. having a wide and open-ended range of applications). This makes it difficult to apply many known risk assessment techniques \citep{koessler2023}. (3) There will likely be significant disagreement about what constitutes an adequate safety case. Experts already disagree about how much assurance existing techniques provide. This disagreement will likely grow for future systems. As such, there is a need to invest in methodology for frontier AI safety cases and to work towards consensus on the ``bar'' a safety case must meet. This is not only necessary for developers in producing high-quality safety cases, but also for reviewers to assess safety cases consistently. Developers are particularly well-placed to address this challenge, but governments can also conduct and coordinate research \citep{irving2024}.

However, developers should not delay producing safety cases until this challenge has been resolved. On the contrary, writing safety cases will likely be essential to make progress on methodology. Yet, decision-makers should be aware that early-stage safety cases will likely be somewhat experimental. They should not rely on safety cases until methodologies are more robust. Regulators specifically may also want to delay using safety cases until clearer expectations and standards can be communicated to developers. At the same time, developing standards will be an iterative process and regulators should expect that they will need updating.

The second technical challenge is to develop the safeguards necessary to assure the safety of advanced future systems. At some point, developers may create frontier AI systems with significant dangerous capabilities.
As discussed in \Cref{components-of-a-safety-case}, we do not yet know how to reliably assure the safety of such systems. So far, safeguards such as harm refusal have proved easy to circumvent via techniques like jailbreaking \citep{chao2023}. Safeguards may become even less robust in the future if systems deliberately act to subvert them \citep{carlsmith2023,hagendorff2024,hubinger2024,jarviniemi2024,pacchiardi2023,park2023,weij2024}. As such, producing safety cases for advanced future systems may require significant breakthroughs in the science of AI safety. Again, developers are likely best-placed to address this challenge, but governments can play a supporting role.

Decision-makers need not wait until this challenge has been resolved before using safety cases to inform decisions. Sufficiently advanced systems should not be deployed until we can be reasonably confident they do not pose unacceptable risk, that is, until it is possible to produce an adequate safety case.\footnote{In theory, there might be cases in which one can be reasonably confident that a system does not pose unacceptable risk, yet it is not possible to produce an adequate safety case for that system. For example, there may be good reasons to believe that a system is safe enough that cannot be easily written down or communicated. However, it is hard to imagine any such cases in practice.} As such, resolving this challenge is not a prerequisite for using safety cases in decision-making, but rather a prerequisite for deploying systems with sufficiently dangerous capabilities.

\subsection{Institutional
challenges}\label{institutional-challenges}

The first institutional challenge is to implement an appropriate structure for reviewing safety cases. This involves setting a clear process, assigning roles and responsibilities, and creating the right incentive structures for the producers and reviewers of safety cases. This challenge applies both to developers and regulators, though they face different challenges. Developers using safety cases internally should ensure that different organizational functions are responsible for creating, reviewing, and stress-testing safety cases. These responsibilities could be assigned and coordinated using the Three Lines Model, which is a popular risk governance framework that helps organizations to allocate different risk management responsibilities \citep{schuett2023,schuett2024-ia}. Regulators using safety cases for decision-making need to appoint or establish a body to review safety cases. This body must be able to receive and store safety cases securely, given that they will likely contain proprietary and sensitive information. Regulators will also need to consider appropriate checks and balances, such as an appeals process or regular auditing of the review body.

The second institutional challenge is for the review body to build sufficient capacity and expertise to effectively review safety cases. The review body would ideally have expertise in frontier AI systems and risk assessment, capacity to review safety cases in depth, information gathering powers, and model access \citep{Bucknall2023-fi}. For developers using safety cases internally, it will be a challenge to balance independence with capacity and expertise. More independent potential reviewers, such as an ethics board \citep{schuett2024}, may have less time and less hands-on expertise with frontier AI models relative to less independent reviewers, such as an internal safety team. For regulators using safety cases in decision-making, it may be a challenge to attract sufficient expertise. This may require funding to pay higher-than-usual salaries to attract private sector talent \citep{hopkins2012}.

While building an effective review system is not a prerequisite for using safety cases in decision-making, it will likely be one of the most important factors in ensuring that they are used well. A core lesson from the other industries is that a capable review body is essential for safety cases to achieve their goals, as safety cases may otherwise be significantly affected by errors and confirmation bias \citep{Haddon-Cave2009,hopkins2012,leveson2011,rinehart}.
This may be even more true for frontier AI, since the novelty and complexity of the technology means that safety case review may be more subjective than usual. As such, decision-makers should be wary of over-relying on safety cases until the reviewer has built sufficient capacity. However, building capacity will be an ongoing process and practice reviewing safety cases would help develop this capacity. Developers and regulators should therefore consider putting in place provisional bodies that initially primarily provide advice or feedback on safety cases, before eventually adopting decision-making powers. Review guidelines may also help make review more efficient and effective at spotting common pitfalls. Such guidelines are common in other industries that use safety cases \citep{draganjovicic2009,militaryaviationauthority,ukofficefornuclearregulationonr2014,ulstandards&engagementulse2023}.

The third institutional challenge is to include third parties in writing and reviewing safety cases. Including third-party organizations could help address multiple of the challenges raised above, including capacity constraints, gaps in reviewer expertise, and lack of reviewer independence. Third party actors could play several roles. First, they could provide evidence for the safety case itself, such as a governance audit report \citep{mokander2023} Second, they could be consulted by the safety case reviewer on specific questions, such as reproducing a model evaluation or reviewing a risk analysis in their domain of expertise. Third, they could review the entire safety case. For example, a developer could commission an independent safety case review to share with relevant decision-makers, such as the board. As another example, a regulator could receive both a safety case from the developer and a “risk case” or red-team of the safety case produced by a third party to help reduce confirmation bias \citep{clymer2024}. 

However, third-party involvement is not a prerequisite for using safety cases in decision-making and need not delay the adoption of safety cases. It is rather a goal to strive for in a mature safety case ecosystem. In addition to determining norms and processes for third party involvement, a key challenge will be to develop the ecosystem in the first place \citep{birhane2024,ukcentrefordataethicsandinnovationcdei2021}. Regulators can support the creation of such an ecosystem (e.g. via funding or other financial incentives).
\newpage

\section{Policy recommendations}\label{policy-recommendations}

\subsection*{Recommendations for developers:}

\begin{itemize}[leftmargin=2em]
\item
  \textbf{Produce safety cases for the next generation of frontier AI systems.} Early safety cases can simply make explicit the arguments that are implicit in current models cards and safety frameworks. As capabilities advance, safety cases may require more rigorous inability arguments (e.g. based on more rigorous risk analysis and externally validated evidence) or other types of arguments (e.g. based on effective safeguards, good governance, or trust in the frontier AI system).

\item
  \textbf{Commit to not deploying future generations of AI systems until a safety case has passed internal review.} Developers could also consider making such a commitment already for the next generation of systems given uncertainty about their capabilities and risks.

\item
  \textbf{Build capacity to internally produce and review safety cases.} For example, assign roles and responsibilities, hire relevant expertise, set up documentation processes, and determine how safety cases will be reviewed.
  
\item
  \textbf{Start the safety case early in development and continue updating it after deployment.} Developers should sketch the safety case and begin to gather documentation already when planning a new training run. Developers should also continue updating the safety case after deployment, both periodically and in response to specific triggers. These processes can be experimental at first as best practices develop.
  
\item
  \textbf{Share safety cases with external stakeholders, especially governments (e.g. AI Safety Institutes).} Developers should also consider sharing (potentially redacted) versions of safety cases with downstream developers, third party research organizations, and the public.
  
\item
  \textbf{Participate in industry-wide development of safety case methodology and best practices.} For example, participate in industry discussions, share best practices with national AI Safety Institutes, and engage in research partnerships.
\end{itemize}

\subsection*{Recommendations for governments:}

\begin{itemize}[leftmargin=2em]

\item
  \textbf{Encourage companies to produce and share safety cases.} For example, secure voluntary commitments and set up the necessary infrastructure such as memorandums of understanding with developers and a platform for securely receiving safety cases. In the longer term, consider offering feedback on safety cases.

\item
  \textbf{Support companies in implementing safety cases.} For example, convene conversations on safety cases and fund or conduct supporting   research such as safety case templates or case studies. By doing so, governments can lower the cost for companies in developing safety cases. In the longer term, engage in a dialogue with industry and third parties to issue guidance on best practices for safety cases.

\item
  \textbf{Contribute to the development of a third-party ecosystem to help with producing and reviewing safety cases}. For example, provide funding or other financial benefits to relevant organizations such as auditors or third-party model evaluators.

\item
  \textbf{Consider using safety cases to assess compliance with existing or future safety requirements on frontier AI developers}. Take into account industry conditions (\Cref{safety-cases-in-regulation}) as well as technical and institutional challenges (\Cref{implementation-challenges}).
  For example, using safety cases is more likely appropriate if capabilities continue to advance rapidly, if basic best practices for safety cases have been established, and if the regulator has the capacity and expertise to effectively review safety cases.

\end{itemize}
\newpage

\section{Conclusion}\label{conclusion}

This paper has argued that safety cases would be a valuable addition to the toolkit of frontier AI governance, both as a part of industry self-regulation and government regulation. Their first main benefit is that they make an explicit, structured argument, which is helpful for checking if risk assessment is comprehensive and valid. Their second main benefit is they provide a flexible way to assess safety, which is important given that the capabilities, safeguards, and our understanding of frontier AI systems are rapidly developing. It is already feasible to produce rudimentary safety cases based on existing safety frameworks, though significant research breakthroughs will likely be needed to produce safety cases for future systems. While technical and institutional challenges remain, developers and regulators should work to address these and move towards safety case adoption.

Several areas of research could help address the challenges of applying safety cases to frontier AI. First, technical research is needed to sketch safety case arguments and understand gaps where additional research must be conducted before a compelling safety case can be produced. Second, technical and policy research is needed on when and how developers should update safety cases after deployment. There may be lessons from industries such as software and autonomous vehicles. Third, policy research is needed on safety case review processes including the role of third parties, accountability mechanisms, and regulator involvement during development and after deployment.

Safety cases are a promising tool that can be applied already now and scaled to much more capable future systems. However, safety cases for frontier AI face unique challenges. The opaque inner workings and open-ended applications of frontier AI systems make it particularly difficult to predict if and how they will cause harm. Advanced future capabilities could make it extremely challenging to produce adequate safety cases. To solve these challenges, there is an urgent need for companies, governments, and civil society to invest in a collaborative research effort.

\section*{Abbreviations}\label{abbreviations}

\renewcommand{\arraystretch}{1.2}
{\small
\begin{tabular}{@{}ll}
CDEI & UK Centre for Data Ethics and Innovation \\
COMAH & The Control of Major Accident Hazards Regulations \\
DSIT & UK Department for Science, Innovation and Technology \\
ERA & European Union Agency for Railways \\
IAPS & Institute for AI Policy and Strategy \\
MAA & Military Aviation Authority \\
MIT & Massachusetts Institute of Technology \\
MoD & UK Ministry of Defence \\
NASA & US National Aeronautics and Space Administration \\
NCSC & UK National Cyber Security Centre \\
NRC & US Nuclear Regulatory Commission \\
ONR & UK Office for Nuclear Regulation \\
RSCR & The Railways Safety Case Regulations \\
SCSC & UK Safety-Critical Systems Club \\
SCSC & Safety-Critical Systems Club \\
UL & Underwriters Laboratories \\
\end{tabular}}

\section*{Acknowledgments}\label{acknowledgments}

We are grateful for valuable feedback and comments from Onni Aarne, Ashwin Acharya, Lukas Berglund, Joe Benton, Alexis Carlier, Stephen Clare, Oscar Delaney, Ben Garfinkel, Arthur Goemans, Ben Hilton, Samuel Hilton, Lewis Ho, Gretchen Krueger, Chris Meserole, Joe O’Brien, Anne le Roux, Rohin Shah, Ben Smith, Akash Wasil, Peter Wildeford, Zoe Williams, Peter Wills, and the cohort of the Centre for the Governance of AI’s Winter Fellowship 2024. All remaining errors are our own.
\newpage

\bibliographystyle{apacite}
\bibliography{ms}

\end{document}